\documentstyle[epsfig,twocolumn,aps]{revtex}
\begin{document}
\title
{\bf Detecting minijet production in $\sqrt{s}$=1.8 TeV
 $p\bar{p}$ collisions with multi-particle 
transverse energy correlation functions
}
\author{ Qing-Jun Liu\thanks{Air-mail:
P.O. Box 918(7) 100039 Beijing, P. R. China.
E-mail: liuqj@hpws5.ihep.ac.cn.}}
\date{\today}
\address
{Institute of High Energy Physics, Academia Sinica\\
P.O. Box 918(4) 100039 Beijing, P. R. China}
\maketitle

\begin{abstract}
    Multi-particle transverse energy correlation(MTEC) functions are 
proposed to study minijet production in high energy $p\bar{p}$ collisions. 
Obtainable with both the D0 and the CDF detector, the high-order 
MTEC functions are shown to be sensitive probes of jet internal structure
as well as promising observables of detecting minijet production
in $\sqrt{s}$=1.8 TeV $p\bar{p}$ collisions. \\
\noindent
PACS number(s): 13.87.Ce, 12.38.Qk, 13.85.Hd, 25.75.Gz\\
\end{abstract}

	That the non-perturbative component of jet
observables is suppressed by some inverse power of the jet energy 
is one of the main properties of the jet observables, ensuing from 
their infrared and collinear safety \cite{GSSW}. Hence observables of 
minijet production initiated by parton with transverse 
energy $E_{T} \leq $ 5 GeV and calculable at the parton level as high 
energy jet, are definitely much more useful than those of high 
energy jet for the study of non-perturbative effects. Therefore detection of 
minijet production, as a first step toward the measurement of minijet
complementary to the measurement of high energy jet, will open another
window of testing QCD, in addition to the study of high energy 
jet \cite{ALB}, the finding of new particles \cite{FASA} and the 
determining of strong coupling constant $\alpha_{s}$ with high 
precision \cite{ADSL}.\par
	Having been found to play an important role to 
explain data of hadronic interactions \cite{CTT,GPWW},
minijets have also been expected \cite{KLL,CT} to be copiously produced
in the pre-equilibrium stage of ultra-relativistic heavy-ion 
collisions and their interactions dominate transverse energy production
in the central rapidity region. However, before convincing evidence of
minijet production is reported, one can not exclude that other 
mechanisms such as expanding quark-gluon plasma \cite{LBM} or soft 
production \cite{MPR} instead of minijet production play the trick. 
Therefore, detection of minijet production is of great importance not 
only to the elementary particle physics but also to the physics of high 
energy nucleus-nucleus collisions.\par
	Unfortunately, minijets can not be identified directly with the 
successful and well developed jet-finding algorithms \cite{JADE,UACD}, 
because the fluctuations of background are large enough to swamp signatures 
of minijet production. For this reason, no convincing evidence of minijet
production has been reported yet. Recently, high-order multi-particle
transverse momentum correlation functions \cite{LIU1,LIU2} are shown
to be promising observables of detecting minijet production in 
$\sqrt{s}$=1.8 TeV $p\bar{p}$ collisions. However, the functions with
the transverse momenta of particles in an event as variables, may be 
obtained with the CDF detector \cite{DCDF} and surely can not be 
obtained with the D0 detector \cite{DDZE}. This is because the D0
detctor has been designed to measure the polar angle $\theta$, the azimuthal
angle $\phi$ and the energy $E$ for most of the particles in an event, while 
the CDF detector to measure also the momentum $P$. 
The purpose of this 
paper is to introduce a new method that can be used for the study
of minijet production with both the D0 and the CDF detector, and 
then to study the possibility of detecting minijet production 
in $\sqrt{s}$=1.8 TeV $p\bar{p}$ collisions.\par
	Originally designed for the study of transverse collective flow 
in heavy-ion collisions at intermediate energies, multi-particle transverse 
momentum correlation functions \cite{LIU1,LIU2} show promise of  
signaling minijet production in high energy $p\bar{p}$ collisions
because the preferential emission pattern \cite{CDDO,OPAL} of a jet 
is similar to that of sideward collective flow in heavy-ion collisions
resulted from the compressed nuclear matter, while that kind of matter
can not be produced in the $\sqrt{s}$=1.8 TeV $p\bar{p}$ collisions,
and also because one can not expect a preferential emission pattern
from the expanding quark-gluon plasma \cite{LBM} or soft 
production \cite{MPR} as stated in Ref. \cite{RXNW}. As for the 
similarity, one is referred to Refs.\cite{GFCK,GGK}, in which it 
was reported that the sphericity 
analysis \cite{GFCK} of jet production in high energy $e^{+}e^{-}$ 
collisions was generalized to analyze data of heavy-ion collisions 
and led to the first observation \cite{GGK} of collective flow in 
intermediate energy heavy-ion collisions. From the similarity it 
can be inferred that this preferential emission pattern, if projected 
on to a plane perpendicular to the beam direction, will induce not 
only multi-particle transverse momentum correlations which are similar
to those observed in heavy-ion collisions \cite{LIU1}, but also 
multi-particle transverse energy correlations.
In this paper, the MTEC functions are proposed to study 
minijet production in high energy $p\bar{p}$ collisions from the point 
of view that energy preferential emission characterizes jet inner 
structure. It is demonstrated that high-order MTEC functions that
can be obtained with both the D0 and the CDF detector are sensitive probes
of jet internal structure as well as promising observables of 
detecting minijet production in $\sqrt{s}$=1.8 TeV $p\bar{p}$
collisions.\par
	The MTEC functions for a sample of collision events can be
calculated following three steps listed below. First of all, 
select $N$-particle sub-events from an event of multiplicity $M$ and
calculate the following variable for each of the sub-events 
\begin{equation}
	U_{N}=\frac
{
|\sum_{i=1}^{N} \vec{ E_{T}^{i}}
|
}
{
 \sum_{i=1}^{N} E_{T}^{i}
} \  \ ,\ \ N \leq M \ \ ,
\end{equation}
where $\vec{E_{T}^{i}}$ is defined as transverse energy vector for the 
$i$th particle in the sub-event, and the $E_{T}^{i}$ is its magnitude. 
With $\hat{\bf x}$ and $\hat{\bf y}$ representing the unit vectors for $x$ 
and $y$ coordinates, the $\vec{E_{T}^{i}}$ for the $i$th particle in the
sub-event is expressed with its polar angle $\theta_{i}$, azimuthal angle
$\phi_{i}$ and energy $E_{i}$ as following:
\begin{equation}
	\vec{ E_{T}^{i}}=E_{i}\ sin \theta_{i} 
        (cos \phi_{i}\ {\bf \hat{x}}+sin \phi_{i}\ {\bf \hat{y}}) \ \ ,
\end{equation}
Secondly, calculate distribution function 
$D(U_{N})$ of sub-events selected from collision events and 
distribution function $B(U_{N})$ of sub-events 
selected from background events. Finally, calculate
$N$-particle transverse energy correlation function $F(U_{N})$ via its 
definition
\begin{equation}
	F(U_{N})=\frac{D(U_{N})}{B(U_{N})} \ \ \ ,
\end{equation}
\par
	For a given collision event, one of its background events is 
obtained in the following way \cite{LIU1,LIU2}: resetting randomly the 
azimuthal angle for each of the $M$ particles between $0$ and $2\pi$ with 
both the polar angle and the energy (or the magnitude of the momentum vector) 
of the particle unchanged. In such constructed background events, there 
are no multi-particle transverse energy correlations caused by minijet 
production.\par
	The sample of events used for the calculation of $F(U_{N})$
usually contains a wide range of multiplicities $M$, and the number 
of entries an event contributes to the $D(U_{N})$ and $B(U_{N})$ 
histograms scales as $g=M!/(M-N)!\ N!$. Consequently, at some values 
of $N$, contributions from events with higher $M$ completely swamp 
those with lower $M$ if all the entries have equal weight. To compensate 
for this, the contribution from an event with multiplicity $M$ is 
weighted by $M/g$, thus ensuring that each event makes
a contribution to the final result that is proportional to the 
multiplicity $M$ of the event \cite{LIU1,LIU2}.\par
	According to Eq. (1), one can calculate $F(U_{N})$ whenever 
$\theta$, $\phi$ and $E$, 
whether or not $P$ for each of the emitted particles 
in an event are measured. Therefore, the MTEC functions 
are obtainable with both of the two detectors the D0 and the CDF 
currently available at FERMILAB.\par
	Next, we present a brief introduction of the Monte-Carlo model
HIJING \cite{XNWG1} used in the following to generate a sample of events for 
our analysis. Combining a QCD inspired model for
jet production with the Lund model\cite{LUND} for jet fragmentation, 
the formulation of HIJING was guided by the Lund FRITIOF\cite{AGN}
and Dual Parton Model\cite{CSTR} for soft processes and,
the successful implementation of perturbative QCD processes in 
PYTHIA\cite{TSM,PYTH} model for hadronic collisions.  
Based on the assumption of independent production of multiple 
minijets, the QCD inspired model determines the number of minijets
per nucleon-nucleon collisions. For each hard or semihard interaction 
the kinetic variables of the scattered partons are determined by 
calling PYTHIA\cite{PYTH} subroutines. The scheme for the 
accompanying soft interactions is similar to FRITIOF model\cite{AGN}.
Fragmentation subroutine of JETSET \cite{JETSET} is called for hadronization. 
It has been reported \cite{XNWG2} that HIJING can consistently 
reproduce many aspects of multi-particle production in $pp$ and 
$p\bar{p}$ collisions at energies of $\sqrt{s}=20$ GeV
to $\sqrt{s}=1.8$ TeV. Because HIJING code includes  minijet 
production in the above mentioned way, it can be served 
as a theoretical laboratory of testifying the MTEC functions as probes of 
minijet production in high energy $p\bar{p}$ collisions.\par
	The sample of events (SES), with which we study minijet 
production in high energy $p\bar{p}$ collisions are obtained in the 
following way \cite{LIU2}. First, generate a sample of 
minimum biased events of $\sqrt{s}=1.8$ TeV $p\bar{p}$ collisions
using HIJING1.2 code with default parameters. Second, drop off every 
high energy jet event in which at least one high energy jet 
with transverse energy larger than 5 GeV is produced, and 
the remaining events form the SES. In fact, two types of events 
comprise the SES. One type of events (NJPES) has neither 
high energy jet production nor minijet production, and the other 
type of events (MJPES) has minijet 
production but no high energy jet production. 
The experimental counterpart of the SES can be 
obtained following the same steps stated above with the aid of the 
widely used jet-finding algorithms \cite{JADE,UACD} of finding 
high energy jet.\par
 	Our study in this paper is based on $10^{6}$ Monte-Carlo events
of minimum biased $\sqrt{s}=1.8$ TeV $p\bar{p}$ collisions,
from which a SES of about 491320 events is collected via the above 
scheme. Because in the CMS frame of the $p\bar{p}$ collisions 
minijet production and particle production 
in the forward hemisphere ({\it i.e}., $\eta> 0$, $0\leq\phi< 2\pi$) 
are both symmetry to those in the backward hemisphere 
({\it i.e}., $\eta < 0$, $0\leq \phi < 2\pi$), 
and additionally the width of jet profiles is about $1$ in 
pseudorapidity \cite{AFS}, our analysis are focused on charged 
particles with $0\leq \eta \leq 1$, $0\leq \phi < 2\pi$. As a result,
about 438400 events of the SES are at our disposal, among which
about 131102 events constitute the NJPES and the other events 
about 307298 form the MJPES actually used in the following analysis.\par
Fig. 1 shows HIJING1.2 results of the MTEC function $F(U_{N})$ in 
the NJPES of the SES. As shown in Fig. 1, $N$-particle transverse energy 
correlation function $F(U_{N})\simeq 1$ for the NJPES, both with and 
without transverse energy cuts. These results demonstrate that 
transverse energy is uniformly distributed in azimuthal angle in the 
central region $0 \leq \eta \leq 1$ of the NJPES where no jets 
are produced.\par
	Fig. 2 shows HIJING1.2 results of the MTEC function
\begin{figure}
\ifx\undefined\psfig\else
\psfig{figure=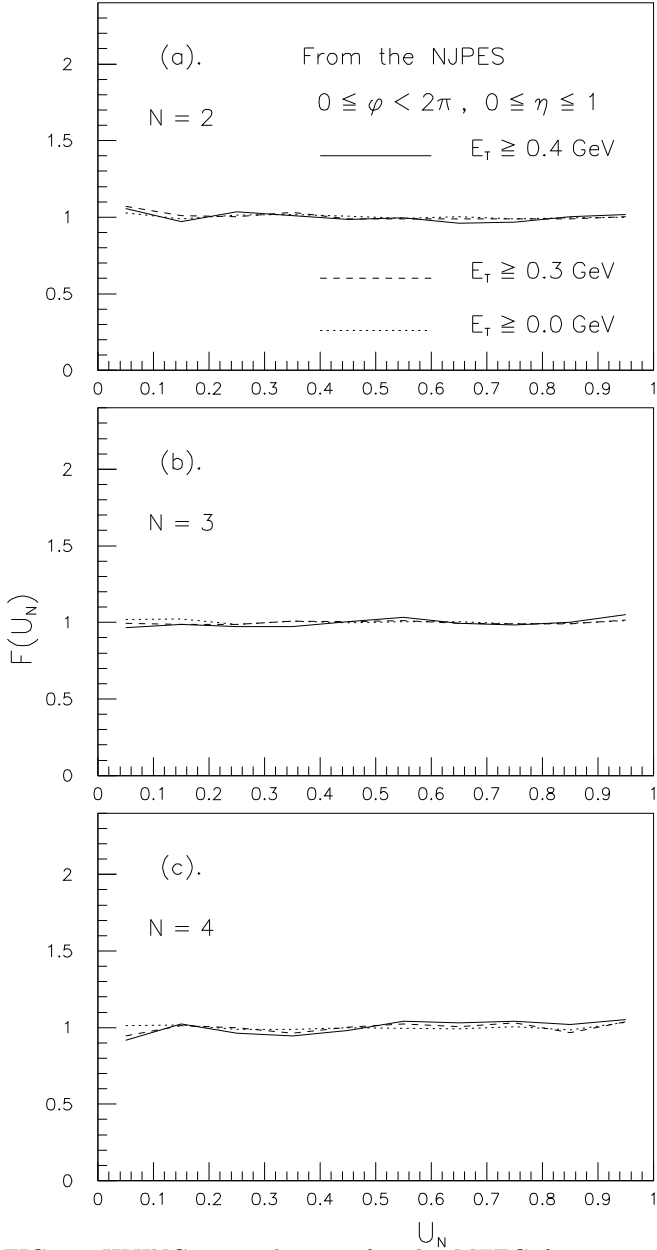, width=\columnwidth}
\caption{HIJING1.2 predictions for the MTEC function $F(U_{N})$ 
in the NJPES of the SES selected from $\sqrt{s}=1.8$ TeV 
minimum biased $p\bar{p}$ collision events.}
\label{fig:eb1}
\fi
\end{figure}
\noindent
$F(U_{N})$ in the SES with and without transverse energy cuts. 
First of all, deviations of $F(U_{N})$ from $1$ appear, {\it i.e}., 
enhancement of $F(U_{N})$ near the region of $U_{N}=1$ and hence 
the systematic suppression of $F(U_{N})$ in other regions of $U_{N}$. 
Comparing with the results shown in Fig. 1, one can conclude that 
these deviations are solely determined by the MJPES of the SES and 
can be inferred to be an indication of minijet production in the 
SES of the minimum biased collision events. The deviations 
characterized by both the enhancement and the suppression 
of $F(U_{N})$ as shown in Fig. 2 result from the preferential 
energy flow induced by minijet production, which definitely 
enhances (suppresses) the probability of finding sub-event with 
high (low) values of $U_{N}$ in the collision events compared to 
that in the background events where no preferential energy flow 
can be expected. Secondly, one can see from Fig. 2 that with the 
increase of sub-event multiplicity $N$, the above mentioned 
deviations of the MTEC function $F(U_{N})$ from 1 become larger. 
A collateral statement is that minijet production can be detected 
more clearly through high-order MTEC functions than through two-particle 
transverse energy correlation function. This feature of MTEC functions 
is originated
\begin{figure}
\ifx\undefined\psfig\else
\psfig{figure=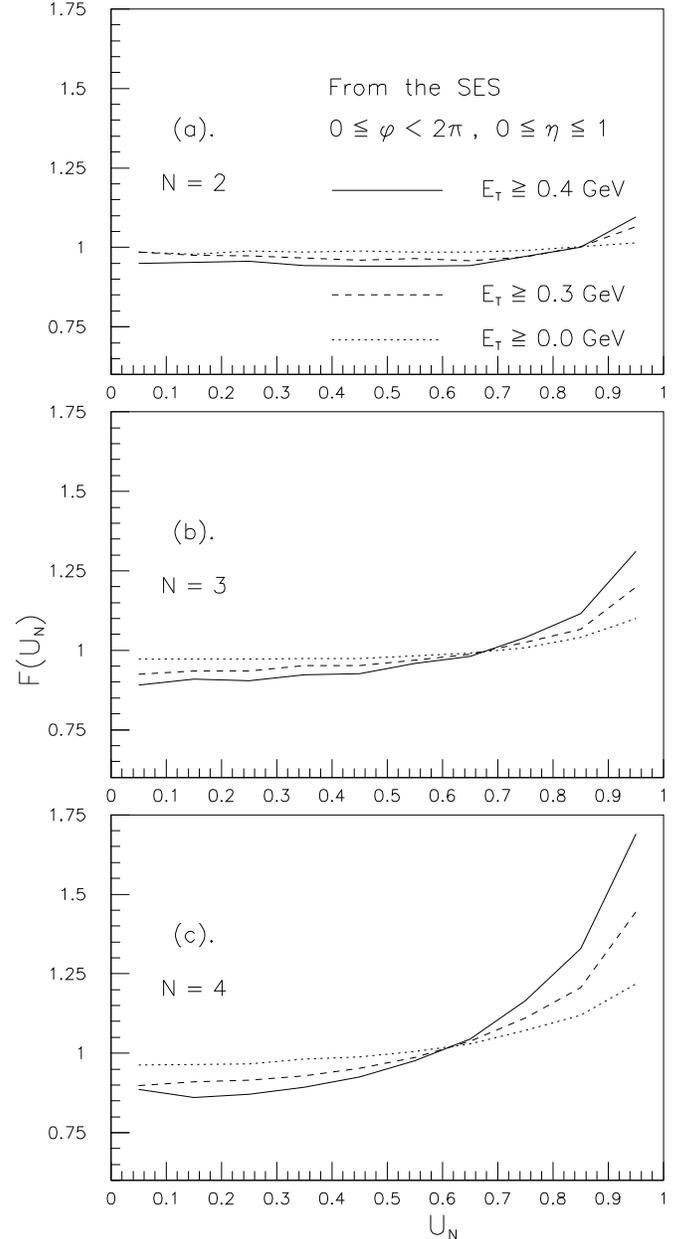, width=\columnwidth}
\caption{HIJING1.2 predictions for the MTEC 
function $F(U_{N})$ in the SES of $\sqrt{s}$=1.8 TeV 
minimum biased $p\bar{p}$ collision events.}
\label{fig:eb2}
\fi
\end{figure}
\noindent
from the collective energy flow of the preferential emission 
pattern of jet production or in other words the collective 
aspact of collinearity property of jet inner structure. 
The two-particle transverse energy correlation function 
reflects this facet of jet property, however not as complete 
as the high-order MTEC functions do. Thirdly, Fig. 2 also 
signifies that $F(U_{N})$ 
calculated from charged particles with a larger transverse 
energy cut displays much more evident enhancement and suppression 
mentioned above than those with a smaller transverse energy cut.
This feature of $F(U_{N})$ shows that particles with higher 
transverse energy are more preferentially emitted 
along the jet axis than those with low transverse 
energy, hence a reflection of another aspact of jet inner structure. 
Therefore, with reasonable 
high transverse energy cuts, high-order MTEC functions can be 
used to study in detail both minijet and high energy jet inner
structure, and can signify validly minijet production 
in $\sqrt{s}=1.8$ TeV $p\bar{p}$ minimum biased collisions.\par
	Now, concluding remarks are given bellow. 
First, new observables, {\it i.e}., the MTEC functions are proposed 
to study minijet production in high energy $p\bar{p}$ collisions. 
As probes of jet inner structure high-order MTEC functions are 
shown to be more sensitive and more complete than 2-particle 
transverse energy correlation function. Second, using the MTEC 
functions, the possibility of detecting minijet production 
in minimum biased $\sqrt{s}$=1.8 TeV $p\bar{p}$ collisions has 
been studied with a Monte-Carlo model. It is demonstrated that 
minijet production can clearly be signaled by enhancement of 
high-order MTEC functions near the region of $U_{N}=1$, 
especially when the functions are calculated with higher transverse 
energy cuts. Compared with the multi-particle transverse 
momentum correlation functions, which can 
be obtained with the CDF detector now at FERMILAB for
1.8 TeV $p\bar{p}$ collisions, the MTEC functions can be obtained
not only with CDF but also with the D0 detector. Therefore,
using the high-order MTEC functions, detection of minijet production 
in minimum biased 1.8 TeV $p\bar{p}$ collisions can be expected 
at FERMILAB in the near future, and a crosscheck on the study of
minijet production with the two mentioned detectors is enabled.\par
The author would like to thank Dr. Xin-Nian Wang for providing
HIJING1.2 code, Professor W.Q. Chao and Professor Y.S. Zhu for
useful discussion. This work was done partly during the author's stay 
in the 7th division of IHEP Beijing and supported by Academia Sinica.


\begin{thebibliography} {40}
\bibitem{GSSW} G. Sterman and S. Weinberg, Phys. Rev. Lett. {\bf 39},
        1436 (1977).
\bibitem{ALB} UA1 Collaboration, C. Albajar {\it et. al}., 
        Nucl. Phys. {\bf B309}, 405 (1988).
\bibitem{FASA} CDF Collaboration., F. Abe {\it et al}., 
	Phys. Rev. Lett. {\bf 74}, 2626 (1995); 
        D0 Collaboration., S. Abachi {\it et al}., 
	Phys. Rev. Lett. {\bf 74}, 2632 (1995).
\bibitem{ADSL} ALEPH Collaboration., D. Buskulic {\it et al}., 
        Phys. Lett. B {\bf 355}, 381 (1995); SLD Collaboration., 
        K. Abe {\it et al}., Phys. Rev. D {\bf 50} 5580 (1994); 
        L3 Collaboration., B. Adeva {\it et al}., 
        Phys. Lett. B {\bf 257}, 469 (1991); DELPHI Collaboration., 
        P. Abreu {\it et al}., Phys. Lett. B {\bf 247}, 
        167 (1990); OPAL Collaboration., M.Z. Akrawy {\it et al}., 
        Phys. Lett. B {\bf 235}, 389 (1990).
\bibitem{CTT} A. Capella and J. Tran Thanh Van, Z. Phys. 
        C {\bf 23}, 165 (1984); L. Durand and H.~Pi, 
        Phys. Rev. Lett. {\bf 58}, 303 (1987); 
	R.C. Hwa, Phys. Rev. D {\bf 37}, 1830 (1988).
\bibitem{GPWW} T.K. Gaisser and F. Halzen, 
        Phys. Rev. Lett. {\bf 54}, 1754 (1985);
	G. Pancheri and Y.N. Srivastava, 
        Phys. Lett. B {\bf 182}, 199 (1986); 
	X.N. Wang, Phys. Rev. D {\bf 43}, 104 (1991).
\bibitem{KLL} K. Kajantie, P.V. Landshoff and J. Lindfors, 
	Phys. Rev. Lett. {\bf 59}, 2572 (1987); J.P. Blaizot
        and A.H. Mueller, Nucl. Phys. {\bf B289}, 847 (1987); 
        K.J. Escola, K. Kajantie and 
        J. Lindfors, Nucl. Phys. {\bf B323}, 37 (1989).
\bibitem{CT} G. Calucci and D. Treleani, 
        Phys. Rev. D {\bf 41}, 3367 (1990);
        G. Calucci and D. Treleani, 
        Phys. Rev. D {\bf 44}, 2746 (1991).
\bibitem{LBM} P. Levai, and B. Muller, 
        Phys. Rev. Lett. {\bf 67}, 1519 (1991).
\bibitem{MPR} C. Merino, C. Pajares and J. Ranft, 
        Phys. Lett. B {\bf 276}, 168 (1992).
\bibitem{JADE} JADE Collaboration., W. Bartel {\it et al}., 
        Z. Phys. C {\bf 33}, 23 (1986); 
        JADE Collaboration., S. Bethke {\it et al}., 
        Phys. Lett. B {\bf 213}, 235 (1988); 
        S. Bethke {\it et al}., Nucl. Phys. {\bf B370}, 310 (1992).
\bibitem{UACD} UA1 Collaboration., G. Arnison {\it et al}., 
        Phys. Lett. {\bf 123B}, 115 (1983); 
        CDF Collaboration., F. Abe {\it et al}., 
        Phys. Rev. D {\bf 45}, 1448 (1992).
\bibitem{LIU1} Liu Qingjun {\it et al}., 
        High Ener. Phys. Nucl. Phys. {\bf 17}, 261 (1993).
\bibitem{LIU2} Q.J. Liu, 'Multi-particle transverse correlation functions 
        and minijet production in $\sqrt{s}$=1.8 TeV $p\bar{p}$ collisions 
        in a Monte-Carlo model', Phys. Rev. D. To be published.
\bibitem{DCDF} CDF Collaboration., F. Abe {\it et al}., 
        Nucl. Instrum. Meth. {\bf A271}, 387 (1988).
\bibitem{DDZE} D0 Collaboration., S. Abachi {\it et al}., 
        Nucl. Instrum. Meth. {\bf A338}, 185 (1994). 
\bibitem{CDDO} CDF Collaboration., F. Abe {\it et al}., 
        Phys. Rev. Lett. {\bf 70}, 713 (1993);
	D0 Collaboration., S. Abachi {\it et al}.,
	Phys. Lett. B {\bf 357}, 500 (1995).
\bibitem{OPAL} OPAL Collaboration., P.D. Acton {\it et al}., 
        Z. Phys. C {\bf 58}, 387 (1993); OPAL Collaboration., 
        P.D. Acton {\it et al}., Z. Phys. C {\bf 63}, 197 (1994).
\bibitem{RXNW} X.N. Wang, Phys. Rev. D {\bf 46}, R1990 (1992).
\bibitem{GFCK} M. Gyulassy, K.A. Fraenkel, and H. Stocker, 
        Phys. Lett. {\bf 110B}, 185 (1982); J. Cugnon {\it et al}., 
        Phys. Lett. {\bf 109B}, 167 (1982); and references therein.
\bibitem{GGK} H.A. Gustafsson {\it et al}., 
        Phys. Rev. Lett. {\bf 52}, 1590 (1984).
\bibitem{XNWG1} X.N. Wang and M. Gyulassy, 
        Phys. Rev. D {\bf 44}, 1501 (1991).
\bibitem{LUND} B.~Andersson, G.~Gustafson, G.~Ingelman and T.~Sj\"{o}strand,
        Phys. Rep. {\bf 97}, 31 (1983).
\bibitem{AGN} B. Andersson, G. Gustafson, and B. Nilsson-Almqvist, 
        Nucl. Phys. {\bf B281}, 289 (1987); B. Nilsson-Almqvist and 
        E. Stenlund, Comput. Phys. Commun. {\bf 43}, 387 (1987).
\bibitem{CSTR} A. Capella, U. Sukhatme and J. Tran Thanh Van, 
        Z. Phys. C {\bf 3}, 329 (1980); A. Capella {\it et al}., 
        Phys. Lett. {\bf 81B}, 68 (1979); A. Capella, C. Pajares and 
        A.V. Ramallo, Nucl. Phys. {\bf B241}, 75 (1984);
        J. Ranft, Phys. Rev. D {\bf 37}, 1842 (1988).
\bibitem{TSM} T.~Sj\"{o}strand and M. van Zijl, 
        Phys. Rev. D {\bf 36}, 2019 (1987).
\bibitem{PYTH} H.-U.~Bengtsson and T.~Sj\"{o}strand,
        Comp. Phys. Commun. {\bf 46}, 43 (1987).
\bibitem{JETSET} T.~Sj\"{o}strand, 
        Comput. Phys. Commun. {\bf 39}, 347 (1986);
        T.~Sj\"{o}strand and M.~Bengtsson, 
        Comput. Phys. Commun. {\bf 43}, 367 (1987).
\bibitem{XNWG2} X.N. Wang and  M. Gyulassy, 
        Phys. Rev. D {\bf 45}, 884 (1992).
\bibitem{AFS} AFS Collaboration., T. Akesson {\it et al}., 
        Z. Phys. C {\bf 30}, 27 (1986); and references therein.
\end{thebibliography}
\end{document}